\documentstyle[11pt]{article}
\begin{document}
 \voffset -2.1cm\hoffset -1.8cm
 \baselineskip 1.7pc\parskip 0.4pc
 \setlength{\unitlength}{0.4pc}

\title{Approximated seventh order calculation of vacuum wave function
of 2+1 dimensional SU(2) lattice gauge theory}

\author{ Ping Hui \\
{\small\em Department of Physics, Guangdong institute of Education
, Guangzhou 510303, China}\\
 Xi-Yan Fang\\
{\small\em Department of Physics, Zhongshan University, Guangzhou
510275, China}\\
 Ting-yun Shi\\
{\small\em State Key Laboratory of Magnetic Resonance and Atomic
and
Molecular Physics,}\\
 {\small\em Wuhan Institute of Physics and Mathematics, the Chinese Academy of Sciences,
 Wuhan 430071,China } }
\date{} \maketitle

\vskip 1.5pc
\begin{abstract}
Using the coupled cluster expansion with the random phase
approximation, we calculate the long wavelength vacuum wave
function and the vacuum energy of 2+1 dimensional Hamiltonian
SU(2) lattice gauge theory (LGT) up to the seventh order. The
coefficients $\mu _0$ , $\mu _2$ of the vacuum wave function show
good scaling behavior and convergence in high order calculations.
\end{abstract}

PACS indices:  11.15.Ha, 12.38.Gc

\newpage

\section{Introduction}

Lattice gauge theory was developed into a promising first
principles approach to the nonperturbative aspects of gauge field
systems. Most of our present knowledge about LGT has obtained from
numerical simulations. However, in order to gain more physical
insight into the theory, it is desirable to develop more
analytical methods. Based on Greensite's proposal \cite{gr}, we
developed the coupled cluster expansions for the Hamiltonian LGT
\cite{GCF,GLF,flg}. Although the preliminary research shows that
this scheme is reasonable to study the LGT, it suffers from the
rapid proliferation and non-independent of clusters produced in
the high order expansive calculations.

In Ref.\cite{GCF},  we used the Cayley-Hamilton relations
\begin{eqnarray}
& Tr U=Tr U^+, ~~U^2=UTr U -1~~~{\rm for~~SU(2)},
\label{e1}\\
& 2Tr U^+=(Tr U)^2-Tr U^2, ~~U^3=U^2Tr U -UTr U^+ +1~~{\rm
for~~SU(3)},\label{e2}
\end{eqnarray}
to eliminate redundancies and find the independent cluster bases,
where $U$ is any group element. But those relations are too
complicated to be used in the high order expansions. We have tried
to use the improved Hamiltonian \cite{lgks} with the tadpole
improvement \cite{lm}, so that the physical results could be
obtained in a relative low order expansion. The research results
for 2+1 dimensional U(1) LGT shows that the improvement of
convergence is immaterial comparing with the complex of
calculation brought about by the improved Hamiltonian \cite{fg}.
Recently, we introduced the random phase approximation (RPA) into
the coupled cluster expansions to circumvent the above problems.
the preliminary investigation results were encouraging \cite{hf}.
In this paper, we use this new method to calculate the vacuum wave
function and vacuum energy of 2+1 dimensional SU(2) LGT.

The paper is organized as follows. In Sec. II, we briefly review
the scheme of the coupled cluster expansion with RPA. Section III
is devoted to the calculation of the vacuum wave function. In Sec.
IV, the conclusions and discussions are presented.

\section{Formulation and Approximation}
The Kogut-Susskind Hamiltonian \cite{KS} is
\begin{equation}
H={g^2\over 2a}\left[\sum_l E_l^2-{4\over g^4}\sum_p Tr(U_p)
\right],
\end{equation}
where the index $l$ denotes the links between sites, $a$ is the
lattice spacing, and $U_p$ the plaquette. The vacuum wave function
can be written as \cite{gr}
 \begin{equation}
 \psi_0(U)=e^{R(U)}.
 \end{equation}
where $R(U)$ consists of various Wilson loops or linked clusters
with appropriate symmetries for the state. In the continuum
limit($a\rightarrow 0$, or equivalently $\beta= 4/g^2\rightarrow
\infty$), the long wavelength behavior of the vacuum state can be
approximated by \cite{GCF}
 \begin{equation}
 \psi(A)=exp\left(-\mu_0\int d^2x {\rm tr}F^2(x)-\mu_2\int d^2x {\rm tr}
 [D_iF^2(x)]+\cdots\right).
 \end{equation}
with $F$ being the field strength tensor and $D_i$ the covariant
derivative. The superrenormalizability of the theory in 2+1
dimensions implies that $\mu_0\rightarrow const./e^2$ and
$\mu_2\rightarrow const. /e^6$, where $e$ is the invariant charge
which is related to the dimensionless coupling constant $g$ by
$g^2=e^2a$.

The eigenvalue equation for the vacuum state of $H$ is
\begin{equation}
\sum_{l} \lbrace [E^a_l,[E^a_l, R(U)]] +[E^a_l,R(U)][E^a_l,R(U)]
\rbrace - {4 \over g^4} \sum_{p}{\rm tr}U_p ={2a \over g^2}
\epsilon_0,\label{bze1}
\end{equation}
where $\epsilon_0$ is the vacuum energy. In the coupled cluster
expansion, $R$ is expanded in terms of a set of linearly
independent cluster $G_{n,i}$ with suitable symmetries
\begin{equation}
R(U)=\sum_{n=1}^M R_n=\sum_{n,i}c_{n,i}G_{n,i},
\end{equation}
with $G_{n,i}$ denoting the $i$-th cluster of order $n$, $c_{n,
i}$ being a coefficient to be determined by Eq.(\ref{bze1}), and
$M$ the highest order number in the coupled cluster expansion.
Substituting it into Eq.(\ref{bze1}) and adopting the truncation
scheme of Ref. \cite{GCF,GLF,flg}, we obtain the truncated
eigenvalue equation
\begin{equation}
\sum_{l} \bigg\{ [E_l,[E_l,\sum_{n=1}^M R_n(U)]] +\sum_{n+n' \le
M}[E_l,R_n(U)][E_l,R_{n'}(U)] \bigg\} - {4 \over g^4} \sum_{p}
{\rm tr}U_p ={2a \over g^2} \epsilon_0. \label{bze2}
\end{equation}

According to the symmetry of the vacuum state, the lowest order
term $R_1$ is chosen to be composed of just one cluster, which is
an elementary plaquette
\begin{equation}
R_1=c_{1,1}G_{1,1}=c_{1,1}\sum_p {\rm tr}U_p.\label{2e1}
\end{equation}
The term $[E^a_l,R_n][E^a_l,R_{n'}]$ in Eq.(\ref{bze1}) will
produce the $(n+n')$-th order clusters. For example,
$[E^a_l,R_1][E^a_l,R_1]$ generates the second order clusters
involving two plaquettes. Obviously, clusters with order $(n+n')$
may contain at most $(n+n')$ Wilson loops. But not all clusters
produced in such a way are independent. They may be related to
each other by relation (\ref{e1}). We have to use this relation to
identify and get the independent clusters. When clusters have many
Wilson loops, relation (\ref{e1}) becomes so complicated that it
is impossible to find a set of independent clusters, as pointed in
Sec. I. To avoid this difficult, the RPA is applied in the
expansions \cite{hf}. In RPA, one set of operator in a product of
two sets of operators is replaced by its average value \cite{wo}.
Here, we replace one Wilson loop by its vacuum average value when
a cluster produced by $[E^a_l,R_n][E^a_l,R_{n'}]$ consists of two
Wilson loops. It is easy to prove that any cluster produced in the
above procedure contains only one Wilson loop. Suppose $R_n$ and
$R_{n'}$ are linear combinations of clusters which contain only
one Wilson loop, then, the new clusters produced by
$[E_l,R_n][E_l,R_{n'}]$ will contain at most two loops, and one
loop will be replaced by its vacuum average after applying the
RPA. On the other hand, according to Eq. (\ref{2e1}) the first
order cluster $G_{1,1}$ has only one Wilson loop. Therefore, all
new clusters produced in the calculation consist of only one
Wilson loop after using the RPA.

Because the clusters consist of only one Wilson loop, the
independent cluster bases for the expansions can be obtained
directly. In addition, the number of independent bases of high
order expansions is much smaller than that without using the
random phase approximation, for example, the number of independent
third order clusters is nine in the coupled cluster expansion
\cite{GCF}, while it is two after using the RPA ( see Sec. III ).
Therefore, the calculation is simplified very much.

The vacuum average value of a Wilson loop can be determined by the
Feynman-Hellman theorem. Let $G$ be some Wilson loop and $<G>$ be
its vacuum average.  Defining $W=H 2a/g^2$, we make the following
change \cite{SW}:
\begin{equation}
W\longrightarrow W^G=W+\xi_G G, \label{2e2}
\end{equation}
where $\xi_G$ is a variable and will take zero at last. From
$W^G|\psi_0>=w^G_0|\psi_0>$, we get
\begin{equation}
<G> =\left.{\partial w^G_0\over \partial \xi_G }
~\right|_{\xi_G=0}, \label{2e3}
\end{equation}

\section{Calculation of approximation}
We now present the calculation of the expansion. Because
 \begin{equation}
 [E^a_l,G_{1,1}][E^a_l,G_{1,1}]=-4-2G_{2,1}+G'_1+G'_2,
 \end{equation}
there are three new clusters produced in the second order
calculation, where the three clusters denoted by $G_{2,1}$,
$G'_1$, and $G'_2$ are given in Fig. 1. Two one which consist of
two Wilson loops turn to cluster $G_{1,1}$ times $<G_{1,1}>$ by
RPA. Therefore, there is only one cluster at second order after
applying the PRA, that is $G_{2,1}$, and
 \begin{equation}
 R_2=c_{2,1}G_{2,1}.
 \end{equation}
 Substituting $R_1$ and $R_2$ into Eq. (\ref{bze2}), we obtain
a set of equations about $c_{1,1}$, $c_{2,1}$ and $w_0$ with a
parameter $<G_{1,1}>$. The parameter $<G_{1,1}>$ can be determined
by Eq. (\ref{2e3}). Solving those equations, we get the second
order approximation of vacuum wave function $\psi_0(U)\approx
e^{R_1(U)+R_2(U)}$. The long wavelength coefficients up to the
second order are
 \begin{eqnarray}
 &&\mu_0=[{c_{1,1}\over 2}+2c_{2,1}]g^4, \\
 &&\mu_2=-{c_{2,1}\over 4}g^8.
 \end{eqnarray}

The third order clusters are produced by term
$[E^a_l,R_1][E^a_l,R_2]$, i.e. $c_{1,1}c_{2,1}[E^a_l,G_{1,1}]
[E^a_l,G_{2,1}]$. Since
 \begin{eqnarray}
& [E^a_l,G_{1,1}][E^a_l,G_{2,1}]&=-2G_{3,1}-G_{3,2}+G'_3+{1\over
2}G'_4+{3\over 2}G'_5-3G_{1,1} \nonumber \\
&&\approx -2G_{3,1}-G_{3,2}+3<G_{1,1}>G_{2,1}-3G_{1,1},\label{2e4}
 \end{eqnarray}
we get
 \begin{equation}
R_3=c_{3,1}G_{3,1}+c_{3,2}G_{3,2}.
 \end{equation}
There are only two independent clusters at order 3, while it is
nine in the expansion without RPA \cite{GCF}, so the calculation
is much simpler. In Eq. (\ref{2e4}), when applying RPA to cluster
$G'_3$, $G'_4$ and $G'_5$, we replace the small Wilson loop with
its vacuum average and let the large one remain unchanged as in
Ref. \cite{hf}. Thus, only one vacuum average of cluster emerges
in the third order calculation. This makes the calculation simple
further. On the other hand, the vacuum or exciting states possess
definite correlation lengths. Only when the space occupied by
glueball is covered with the Wilson loops, the calculation is
efficient, so, we replace the small loop with its vacuum average
and preserve the large loop. Submitting $R_1$, $R_2$ and $R_3$
into Eq. \ref{bze2}, and solving the equation, we obtain the third
order approximation of the vacuum wave function.

Higher order calculation can be carried out similarly. We have
done the calculation up to the seventh order. the number of
independent clusters are 1, 1, 2, 6, 14, 44, and 109 at order 1,
2, 3, 4, 5, 6, and 7 respectively. The results for $\mu_0$ and
$\mu_2$ from the fourth order to the seventh order are presented
in Fig. 2. In Fig. 3, the vacuum energy $w_0$ against $\beta$ are
plotted.

\section{Results and discussions}

From Fig.2, we see that the curves of $\mu_0$ (or $\mu_2$) show
good scaling behavior and convergent trend in weak coupling region
$\beta $=5.5 to 9.5. The sixth and seventh order values of $\mu_0$
are coincident in the scaling region, which shows that the
approach is effective and rapidly convergent. From the seventh
order results, we obtain
\begin{eqnarray}
\mu_0 = 2.3,\label{3e1}\\
\mu_2 = -0.3.
\end{eqnarray}
A Monte Carlo measurement was gave by Arisue \cite{ars}
\begin{eqnarray}
\mu_0 =(0.91 \pm 0.02),\nonumber \\
\mu_2 =-(0.19\pm 0.05). \label{3e2}
\end{eqnarray}
The value of $\mu_0$ in Eq. (\ref{3e1}) is larger than Arisue's
value. It is perhaps that there is some systematic error
introduced by the random phase approximation in our procedure. But
it is also probable that the Monte carlo results of Eq.
(\ref{3e2}) was not enough accurate. Because the values in Eq.
(\ref{3e2}) can be reproduced in the low order ( the third order )
expansion without RPA, while in higher order ( the fourth order )
calculation the values became large \cite{GCF}.

In Fig. 3, the curve of the sixth and seventh order results are
almost coincident, which proves the expansion is able to converge
rapidly again. We also give the third order result of the vacuum
energy calculated without RPA for comparing. The third order
values of the vacuum energy with RPA are lower than that without
RPA. Such a case was also true in the case of SU(3) LGT \cite{hf}.
We think it means that the vacuum state obtained by using RPA is
better than that without using RPA, and it may explain the rapid
convergence of the approach.

\centerline{\bf Acknowledgments} The Project was supported by
Natural Science Foundation of Guangdong Province (33446)and by
Appropriative Researching Fund for Professors and
Doctors,Guangdong Institute of Education.

\newpage
\centerline{\bf Figure Captions}
\parindent 0pc

 \vskip 0.1cm
Fig.1 ~The linked clusters used in the expansions up to order 3.

 \vskip 0.1cm
Fig.2 ~$\mu_0$ and $\mu_2$ as a function of $\beta=4/g^2$. The
four curves represents the results from the fourth order to the
seventh order expansion with RPA respectively.

 \vskip 0.1cm
Fig.3 ~The vacuum energy vs $\beta$.  We also give the third order
vacuum energy calculated without using RPA.

 \newpage
 \begin{picture}(80,14)
\put(3,3){\line(1,0){4}}\put(3,3){\line(0,1){4}}
\put(3,7){\line(1,0){4}}\put(7,3){\line(0,1){4}}
\put(3,-1){\makebox(0,0)[l]{$G_{1,1}$}}

 \put(14,3){\line(1,0){8}}\put(14,3){\line(0,1){4}}
 \put(14,7){\line(1,0){8}}\put(22,3){\line(0,1){4}}
 \put(16,-1){\makebox(0,0)[l]{$G_{2,1}$}}

 \put(29,3){\line(1,0){4}}\put(29,3){\line(0,1){4}}
 \put(29,7){\line(1,0){4}}\put(33,3){\line(0,1){4}}
 \put(33.6,3){\line(1,0){4}}\put(33.6,3){\line(0,1){4}}
 \put(33.6,7){\line(1,0){4}}\put(37.6,3){\line(0,1){4}}
 \put(31,-1){\makebox(0,0)[l]{$G'_1$}}

 \put(44,3){\line(1,0){4}}\put(44,3){\line(0,1){4}}
 \put(44,7){\line(1,0){4}}\put(48,3){\line(0,1){4}}
\put(44.5,3.5){\line(1,0){3}}\put(44.5,3.5){\line(0,1){3}}
\put(44.5,6.5){\line(1,0){3}}\put(47.5,3.5){\line(0,1){3}}
\put(46,-1){\makebox(0,0)[l]{$G'_2$}}

 \put(3,-16){\line(1,0){8}}\put(3,-16){\line(0,1){8}}
 \put(3,-8){\line(1,0){4}}\put(7,-12){\line(0,1){4}}
\put(7,-12){\line(1,0){4}}\put(11,-16){\line(0,1){4}}
\put(5,-20){\makebox(0,0)[l]{$G_{3,1}$}}

 \put(18,-16){\line(1,0){12}}\put(18,-16){\line(0,1){4}}
 \put(18,-12){\line(1,0){12}}\put(30,-16){\line(0,1){4}}
 \put(20,-20){\makebox(0,0)[l]{$G_{3,2}$}}

\put(36,-12){\line(1,0){4}}\put(36,-12){\line(0,1){4}}
\put(36,-8){\line(1,0){4}}\put(40,-12){\line(0,1){4}}
 \put(36,-16.6){\line(1,0){8}}\put(36,-16.6){\line(0,1){4}}
 \put(36,-12.6){\line(1,0){8}}\put(44,-16.6){\line(0,1){4}}
 \put(37.5,-20){\makebox(0,0)[l]{$G'_3$}}

 \put(50.4,-16){\line(1,0){4}}\put(50.4,-16){\line(0,1){4}}
\put(50.4,-12){\line(1,0){4}}\put(54.4,-16){\line(0,1){4}}
 \put(55,-16){\line(1,0){8}}\put(55,-16){\line(0,1){4}}
 \put(55,-12){\line(1,0){8}}\put(63,-16){\line(0,1){4}}
 \put(55,-20){\makebox(0,0)[l]{$G'_4$}}

 \put(69.5,-16){\line(1,0){3}}\put(69.5,-16){\line(0,1){3}}
\put(69.5,-13){\line(1,0){3}}\put(72.5,-16){\line(0,1){3}}
 \put(69,-16.5){\line(1,0){8}}\put(69,-16.5){\line(0,1){4}}
 \put(69,-12.5){\line(1,0){8}}\put(77,-16.5){\line(0,1){4}}
 \put(70,-20){\makebox(0,0)[l]{$G'_5$}}
\put(4,-28){\makebox(0,0)[l]{Fig. 1 ~~The linked clusters used in
the expansions up to order 3.}}
 \end{picture}

\end{document}